\documentclass{webofc}
\usepackage[varg]{txfonts}   % Web of Conferences font
\begin{document}
\title{Constraints of the Low-$x$ Structure of Protons and Nuclei}

\author{\firstname{Thomas} \lastname{Peitzmann}\inst{1}\fnsep\thanks{\email{t.peitzmann@uu.nl}} }

\institute{Institute for Subatomic Physics, Utrecht University, The Netherlands }

\abstract{%
I review recent developments in the study of the low-$x$ partonic content of protons and nuclei, with a focus on the latter, as one expects possible deviations from linear QCD evolution to be most pronounced in that case. I give examples of recent theoretical descriptions of HERA measurements with a focus on the role of BFKL evolution. 
I then concentrate on the status and assumptions of nuclear PDFs and the possibility to use forward particle production at the LHC as further constraint, in particular measurements of open charm and the potential of electromagnetic probes. }
\maketitle
\section{Introduction}
\label{intro}
Recent high-energy physics experiments, most notably at the LHC, are providing high precision measurements, which have to be confronted with similarly precise theoretical calculations. For QCD processes, important quantitative ingredients to theoretical calculations, the parton distribution functions, carry considerable uncertainties, which in turn limits the predictive power of theory. In particular, gluon PDFs are only very poorly constrained. Advances on theoretical and experimental sides are needed to improve the status.

In addition, it has been realised that the increasing parton density for small $x$ at moderate $Q^2$ should become problematic for a description via a purely linear QCD evolution, via DGLAP or BFKL equations. In fact, current frameworks describing initial state parton dynamics rely entirely on DGLAP evolution, so even BFKL has not been well established as a theoretical tool. At high density, partons should start to overlap, leading to recombination processes. The evolution would have to be described by non-linear equations, the JIMWLK equations, or approximately by the BK equations. The corresponding non-linear processes should lead to a saturation of the parton density at small $x$ below a characteristic momentum scale, the \textit{saturation scale} $Q_s$. As for small $x$ gluons are the most important degree of freedom, this phenomenon is usually termed \textit{gluon saturation}.
The transverse gluon density in a nucleus should increase with $A^{1/3}$, so one expects gluon saturation effects to be stronger in heavy nuclei. This is reflected in the expected dependence of the saturation scale on $x$ and $A$:
\begin{equation}
Q_s^2(x) \approx \frac{\alpha_s}{\pi R^2} x G(x,Q^2) \propto A^{1/3} \cdot x^{-\lambda},
\end{equation}
with $\lambda \approx 0.3$. A potential proof of gluon saturation would be related to the observation of strong nuclear modification of the gluon density and a deviation from linear QCD evolution.
 
\section{Proton PDFs}
\label{sec-1}
Most of the recent work on proton PDFs is about improved theoretical predictions. As this is not a comprehensive review, I will just give two specific examples, both related to BFKL evolution.

\begin{figure}[h]
\centering
\includegraphics[width=0.9\textwidth]{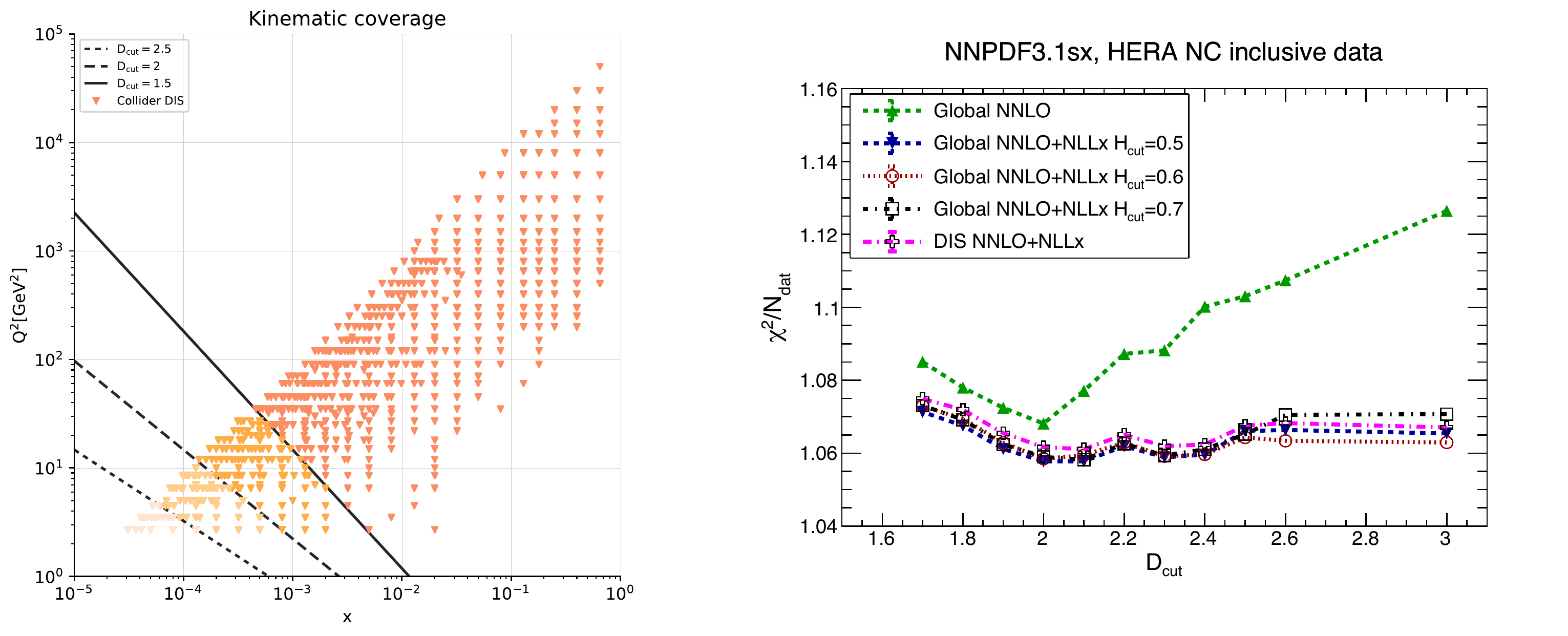}
\caption{Left: Kinematic coverage of DIS data as a function of $x$ and $Q^2$ compared to cut values used in the NNPDF fits. Right: $\chi^2$ per degree of freedom for different theory implementations as a function of the value of the kinematic cut used. (From \cite{ball-2018}.)}
\label{fig-bfkl1}       % Give a unique label
\end{figure}
\vspace{-2mm}

In \cite{ball-2018}  the authors have performed a systematic comparison of different theoretical approaches to the description of DIS data. They have implemented BFKL effects in the NNPDF framework and have performed global fits of HERA data using either fixed-order calculations (NNLO) or calculations using also small-$x$ resummation (NNLO+NLLx). Fits have been performed in different kinematic regions, which is controlled via a parameter $D_{\mathrm{cut}}$ as shown in the left panel of Fig.~\ref{fig-bfkl1}. Small values of this parameter exclude a part of the low-$x$ and low-$Q^2$ region, so that a fixed-order calculation should be sufficient to describe the data, while large values include more of the data where the small-$x$ resummation should become important. The behaviour of the fit quality is displayed in the right panel of Fig.~\ref{fig-bfkl1}, which shows the $\chi^2$ per degree of freedom as a function of $D_{\mathrm{cut}}$ for the different calculations. For small values of $D_{\mathrm{cut}}$, all calculations describe the data equally well, i.e. small-$x$ resummation is not really needed, while for large values of of $D_{\mathrm{cut}}$ the fixed-order calculation shows a significantly larger  $\chi^2$. The behaviour for inclusion of the small $x$, small $Q^2$ data strongly suggest that BFKL evolution is necessary to describe the data, and that this particular kinematic region is sensitive to it.
\begin{figure}[h]
\centering
\includegraphics[width=0.6\textwidth]{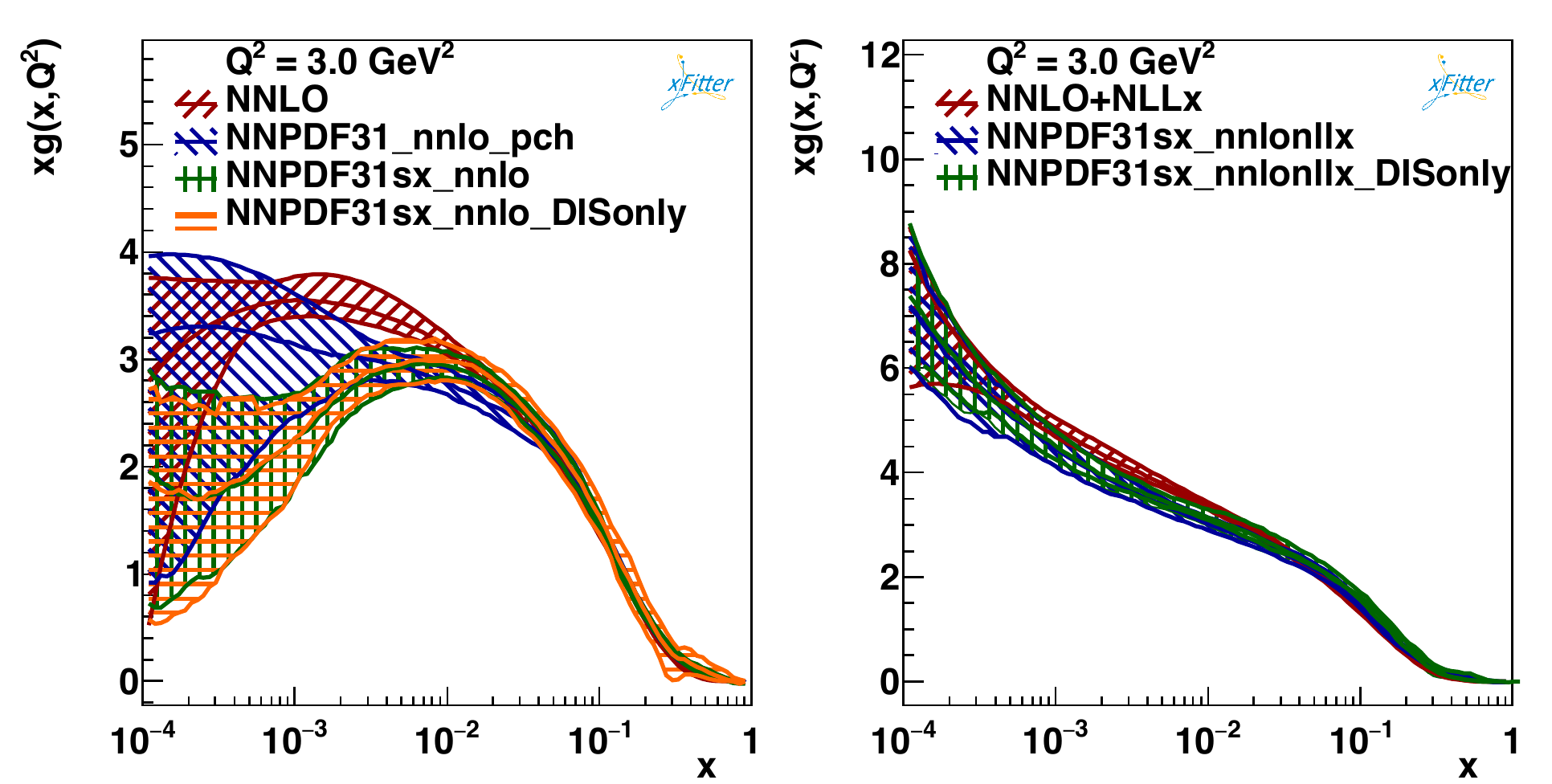}
\caption{Gluon PDFs resulting from fits of different theoretical descriptions (HERAPDF and NNPDF) to HERA DIS data. Left: Fits using fixed-order calculations. Right: Fits including $1/x$-resummation. (From \cite{abdol-2018}.)}
\label{fig-bfkl2}       % Give a unique label
\end{figure}
%\vspace{-2mm}

A related analysis is performed in \cite{abdol-2018} implementing both conventional fixed-order approaches and also approaches including BFKL evolution in the xFitter program, and similar conclusions are found. As an example illustrating their findings, Fig.~\ref{fig-bfkl2} shows gluon PDFs obtained from different fits using the HERAPDFs and as alternative the NNPDFs. The left panel shows results using NNLO results - here it is apparent that there is a strong tension between the different approaches, particularly at small $x$. This is remedied when the fits include BFKL evolution as shown in the right panel, where the different approaches are consistent with each other. These findings demonstrate clearly the necessity of incorporating BFKL dynamics also in the context of PDF fits. 

No clear sign of saturation has been observed in the behaviour of parton densities in protons. To observe potential deviations from linear evolution one apparently needs additional constraints from measurements at still smaller $x$, which are currently not available.

\section{Parton distributions in nuclei}
\label{sec-2}
\begin{figure}[h]
\centering
\includegraphics[width=0.45\textwidth]{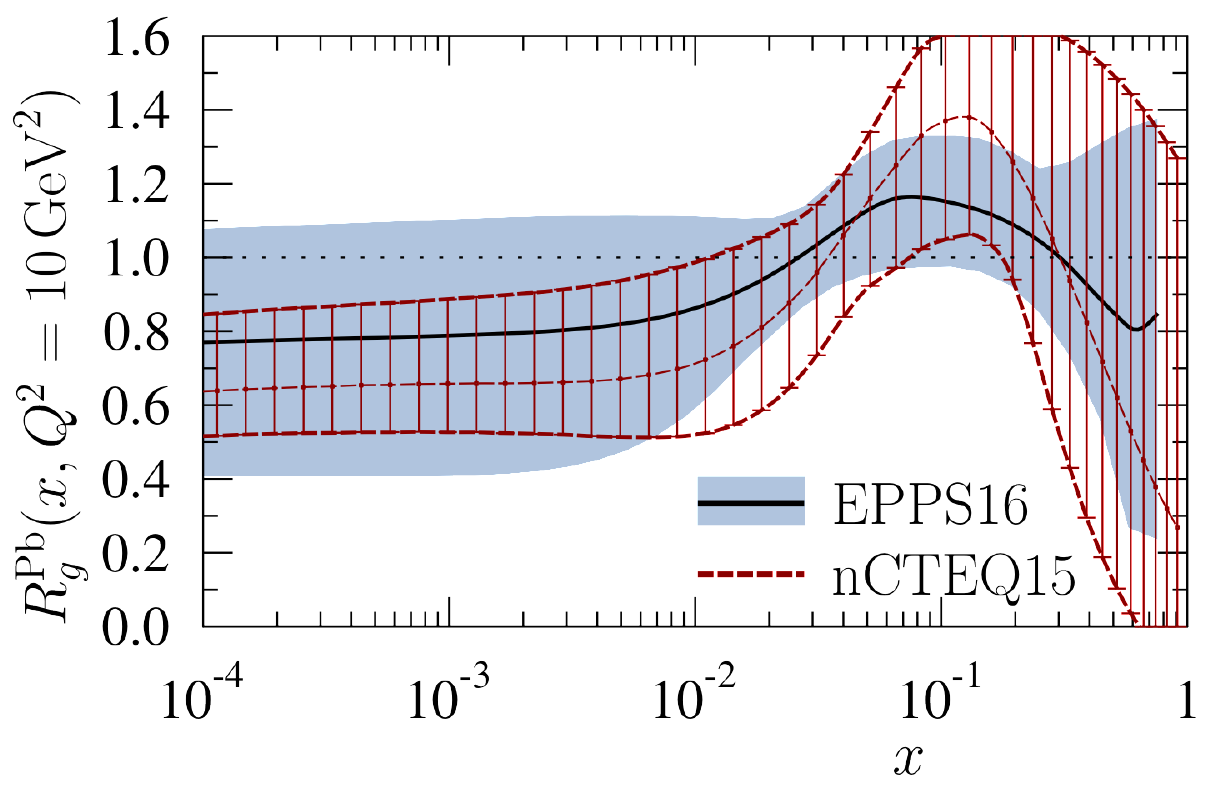}
\includegraphics[width=0.4\textwidth]{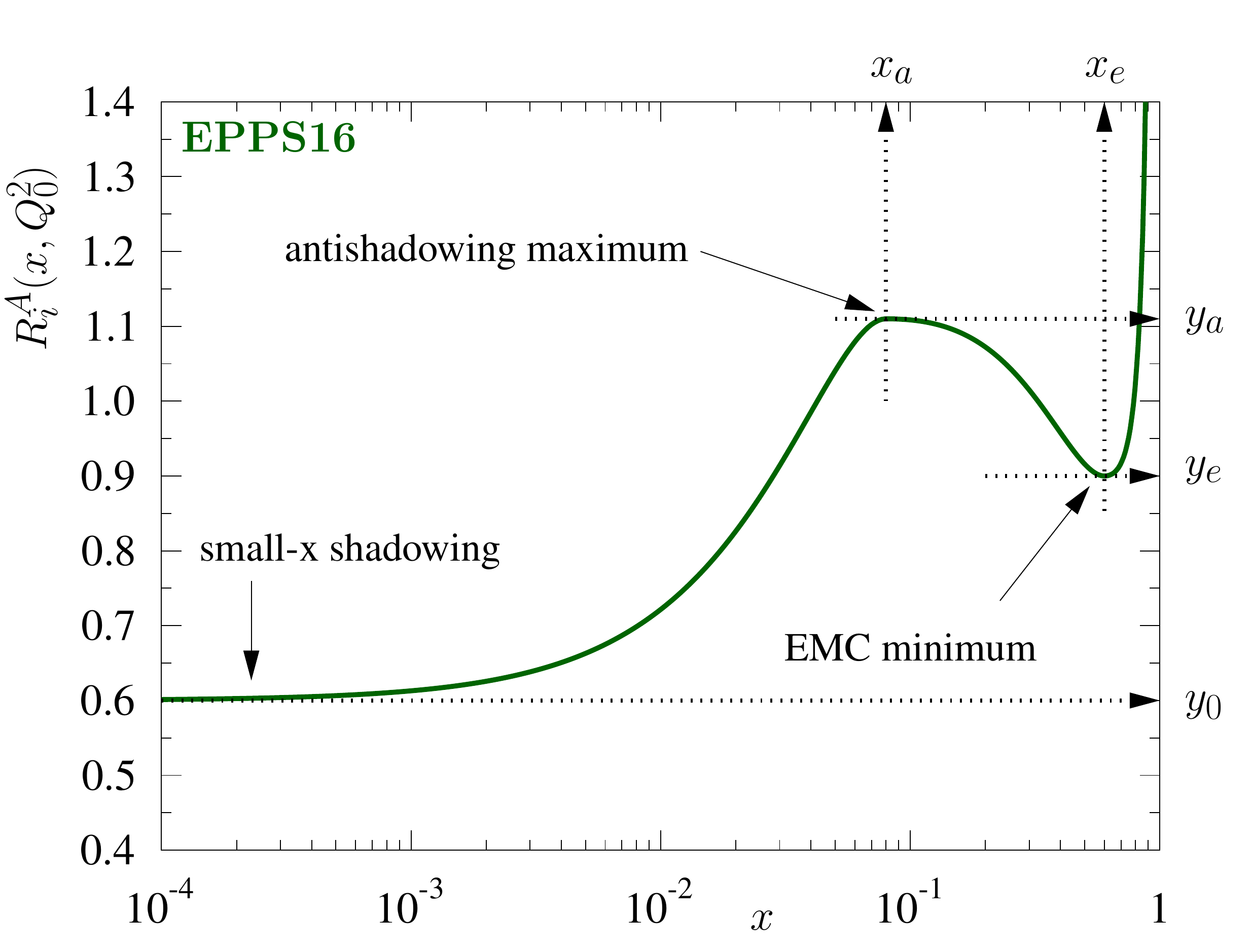}
\caption{Left: Nuclear modification factors $R_g^{\mathrm{Pb}}$ of gluon PDFs in Pb at $Q^2 = 10\, \mathrm{GeV}^2$ as obtained in the EPPS16 and nCTEQ15 frameworks. Right:  Analytic shape of $R_g^{\mathrm{A}}$ as used in EPPS16. (From \cite{epps16}.)}
\label{fig-nPDF1}       % Give a unique label
\end{figure}

Parton distributions are still much less known in nuclei. Very little nuclear DIS data is available, and the kinematic reach is very limited. At the same time, knowledge of the parton densities would be extremely interesting here, as saturation effects should be much more prominent. The current state of theoretical descriptions can be illustrated by the nuclear modification factors of the PDFs, which parameterise the difference of the parton distributions in nucleons inside a nucleus relative to the free nucleon.  The left panel of Fig.~\ref{fig-nPDF1} shows examples of the gluon nuclear modification factor from state-of-the-art nuclear PDF sets.

\vspace{-2mm}

Both nuclear PDFs show qualitative features that are expected, in particular a so-called antishadowing maximum at moderate $x$ and a value smaller than one for small $x$, which is usually attributed to shadowing. Both sets show large uncertainties. We are most interested in the features of these distributions at small $x$, and it is notable that both descriptions show very little to no $x$ dependence of $R_g^{\mathrm{Pb}}$ there. This is related to the analytical shapes of functions allowed for the description of $R_g^{\mathrm{A}}$. This is illustrated by the right hand panel of the figure, which displays the parameterisation used in EPPS16 nPDFs. The function levels off to a constant value at small $x$, which is an invariant feature of the parameterisation used. As the shape of the function is similar for nCTEQ15, we have to assume a similar feature in their parameterisation. The corresponding fit shows smaller uncertainties with usage of the same data, which may be related to a different treatment of flavour dependence or even less flexibility in the shape of the nCTEQ15 parameterisation. Stronger modifications of the PDFs appearing at very small $x$ cannot be described by these functions, which is a limitation that biases the uncertainties obtained. We will revisit this issue again below.

As nuclear DIS data are scarce, one can use particle production at the LHC to constrain the PDFs. In $pA$ collisions one could thus obtain important constraints on nuclear PDFs. Using parton level kinematics at leading order, a measurement of particles with a given transverse momentum $p_T$ at rapidity $y$ in collisions with a CM energy of $\sqrt{s}$ would be sensitive to Bjorken $x$ values as low as:
\begin{equation}
x = \frac{2 p_T}{\sqrt{s}} \exp (-y)
\label{eq-x}
\end{equation} 
\begin{figure}[h]
\centering
\includegraphics[width=0.5\textwidth]{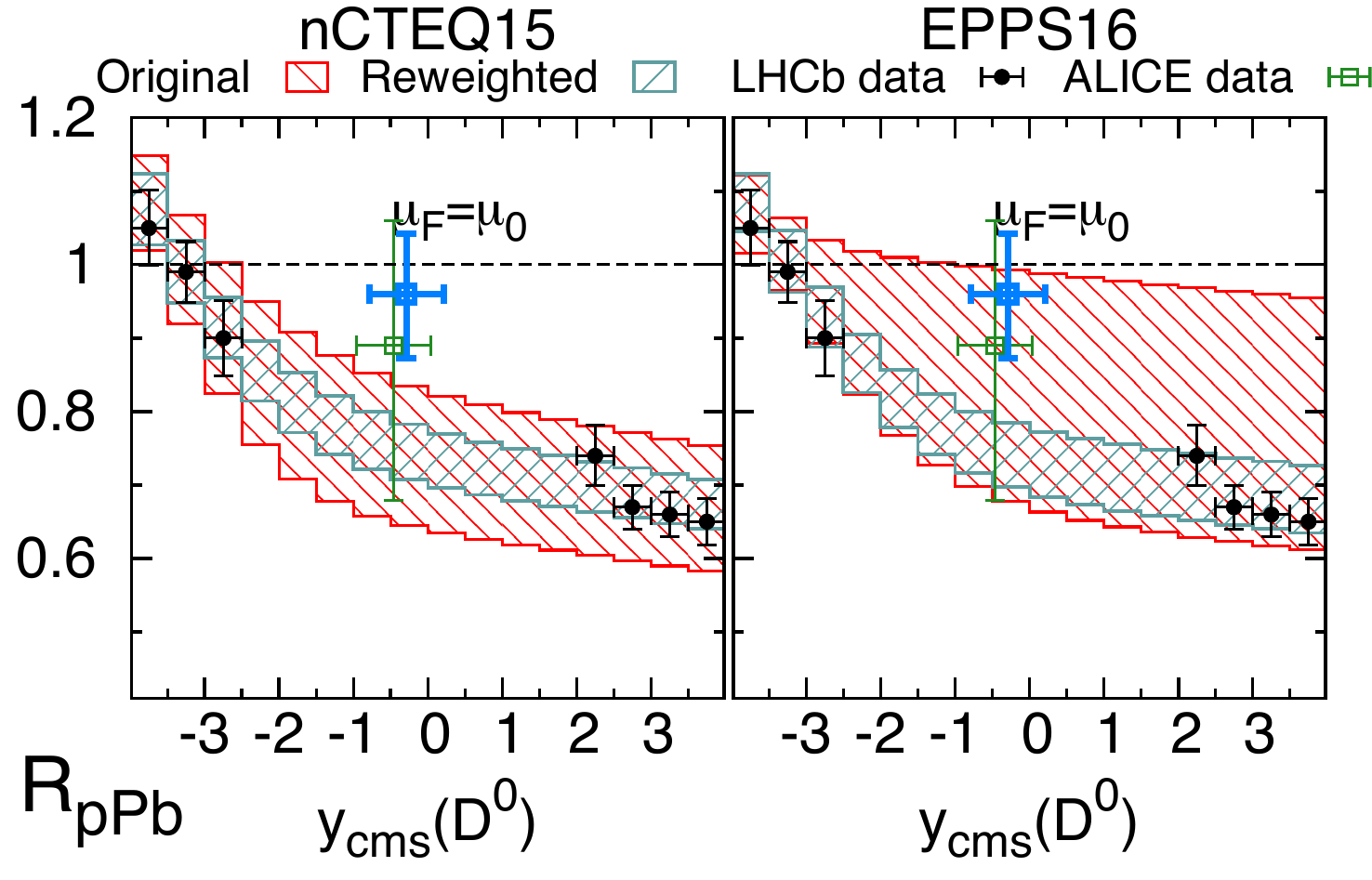}
\caption{Nuclear modification factor for $D^0$ production in p--Pb collisions at the LHC as a function of rapidity. Measurements from LHCb and ALICE are compared to refits of nuclear PDF sets nCTEQ15 (left) and EPPS16 (right) for one choice of scales. Bands show the original uncertainties (red) and those after the reweighting including the charm data (grey). The figure is reproduced from \cite{kusina-2018}, in addition, recent data from ALICE \cite{alice-charm}, which have not been used in the reweighting, are added as blue symbols.}
\label{fig-kusina-charm}       % Give a unique label
\end{figure}
\vspace{-2mm}

So far, measurements of electroweak probes or jets, which are preferable as the production processes are well understood, are only available at midrapidity and relatively high $p_T$, so these would constrain rather high values of $x$ only. Forward measurements at not too high $p_T$ have strong advantages. Considering the lack of forward EW probes, open charm measurements are the most promising among existing measurements at the LHC, as charm production should proceed via hard scattering processes. First attempts have been made to include charm production into refits of nPDFs. The results of one such analysis \cite{kusina-2018} are shown in Fig.~\ref{fig-kusina-charm}. Data from LHCb and preliminary data from ALICE have been used in a reweighting procedure, both with nCTEQ15 and EPPS16. At backward rapidity, the nuclear modification factor $R_{\mathrm{pPb}}$ is consistent with one, i.e. no modification, while it shows a strong suppression at forward rapidity. The calculations follow nicely the LHCb results, and the inclusion of those data reduces the apparent uncertainty considerably. 

Interestingly, $R_{\mathrm{pPb}}$ shows very little variation between midrapidity and forward rapidity. While this appeared to be consistent with data in view of the preliminary ALICE results with their large uncertainties, there is some tension between the calculation and the more recent published ALICE results at midrapidity \cite{alice-charm}, which are added to the figure as blue symbols. The sensitivity range in $x$ of a measurement depends on rapidity (see equation \ref{eq-x}), so it seems likely that the lack of rapidity dependence in $R_{\mathrm{pPb}}$ can be traced to a lack of $x$-dependence in the nuclear modification of the PDFs, i.e. $R_g^{\mathrm{Pb}}$. This is a strong hint that the assumptions underlying the functional shapes used in the PDF sets should be revisited.
\vspace{-2mm}

\begin{figure}[h]
\centering
\includegraphics[width=0.35\textwidth]{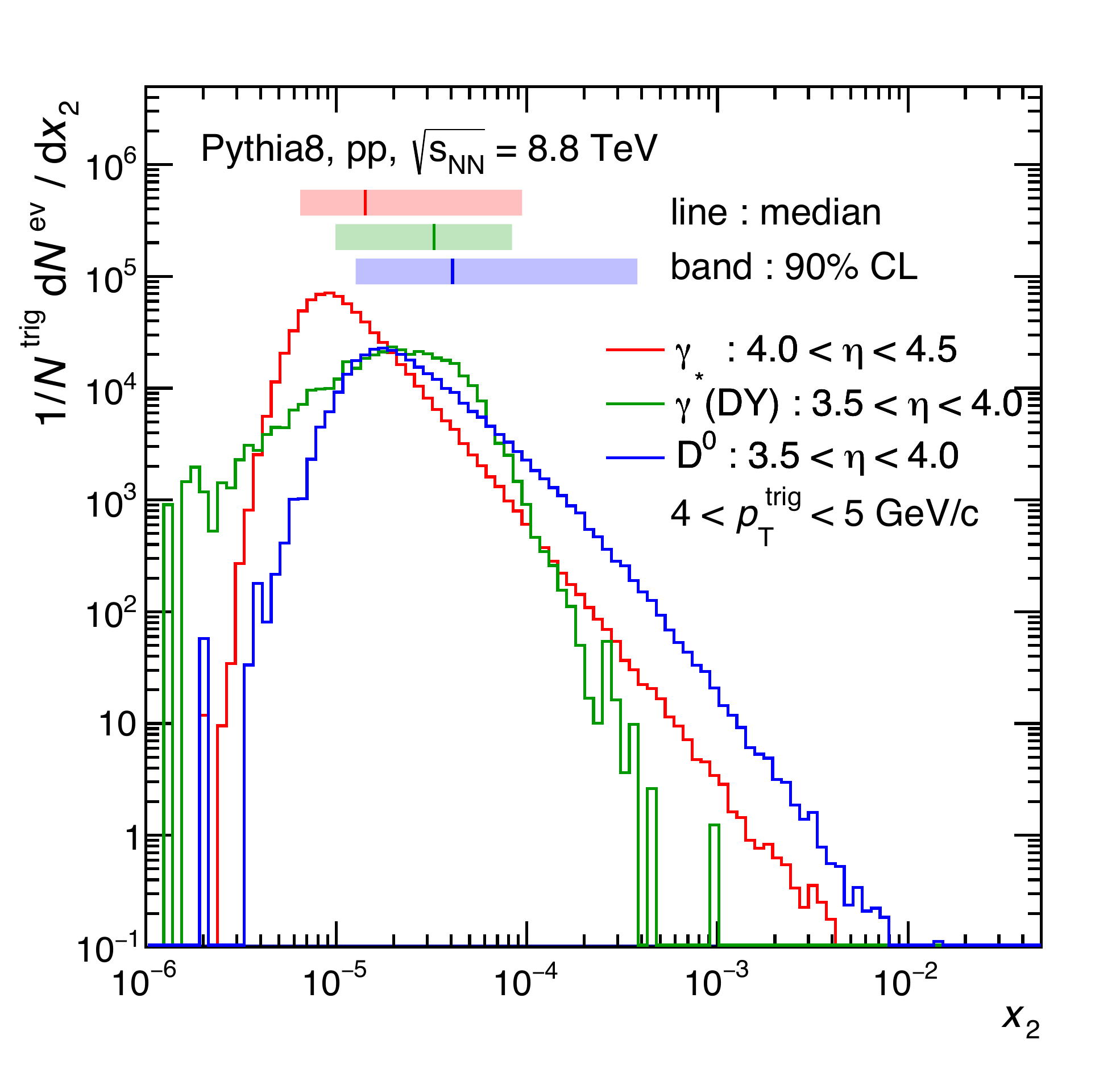}
\includegraphics[width=0.38\textwidth]{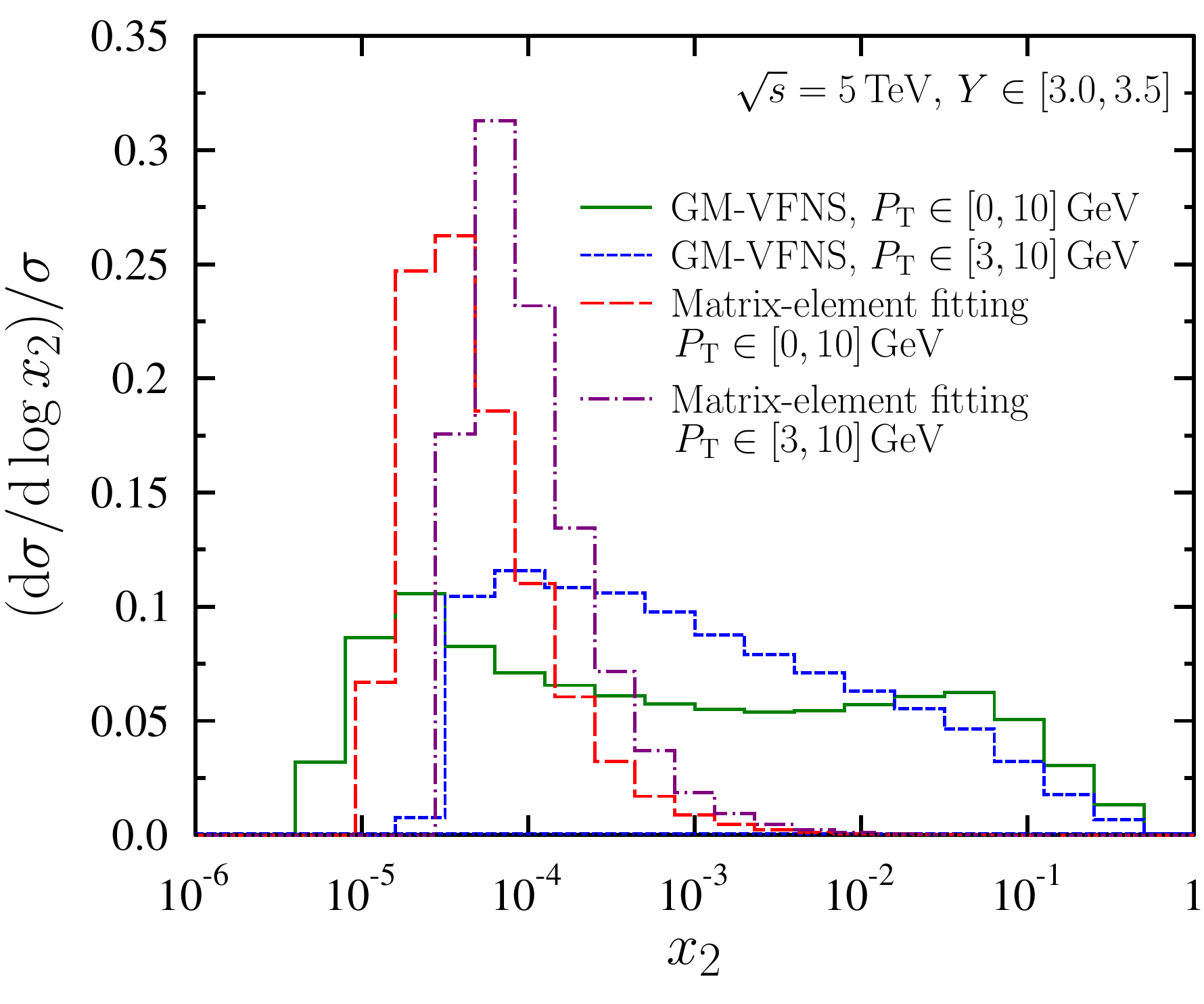}
\caption{Distributions of $x_2$ probed in forward production at LHC. Left: Distributions for DY pairs and $D^0$ mesons in LHCb and for real photons in the  proposed forward photon detector (FoCal) in ALICE, from PYTHIA8 calculations. Right:  Distributions for $D^0$ production using different theoretical approximations (from \cite{epps-charm}).}
\label{fig-xdist}       % Give a unique label
\end{figure}
\vspace{-2mm}
When nuclear PDF parameterisations allow for more significant $x$ dependence, the different sensitivity of given probes to specific $x$-ranges become more important. In this context, it is interesting to compare the sensitivity of different probes - this has been done in Fig.~\ref{fig-xdist}. The left panel shows the distributions for different probes of similar $p_T$ as obtained from PYTHIA simulations. All probes are sensitive to values of $x < 10^{-5}$, but the distributions have significant tails towards very large $x$. For measurements possible within LHCb, DY pairs would have a strong advantage from their $x$-sensitivity, however, they are sensitive to gluons only at next-to-leading order, and, likely more important, their cross section is extremely small making a measurement very challenging. A measurement of real photons, as it should become possible with the proposed forward calorimeter (FoCal) in ALICE \cite{Peitzmann:2018kie,FoCal-LoI}, would have an $x$-sensitivity similar to DY, but would not suffer from those disadvantages.

The usage of charm measurements for PDF fits may suffer from another complication. The authors of \cite{epps-charm} have shown that a more realistic study can be done using the variable flavour number scheme (VFNS), and the corresponding x-sensitivity distributions show an even much stronger tail towards large $x$, as displayed in the right panel of Fig.~\ref{fig-xdist}. This suggests, that the sensitivity of charm production to very small $x$ may be much more limited than previously thought. The situation is even more complex, when considering the hints for final state modification of charm distributions, as seen in the nuclear modification  \cite{alice-charm} and the values of elliptic flow $v_2$ 
\cite{cms-charm-v2}.

\begin{figure}[h]
\centering
\includegraphics[width=0.35\textwidth]{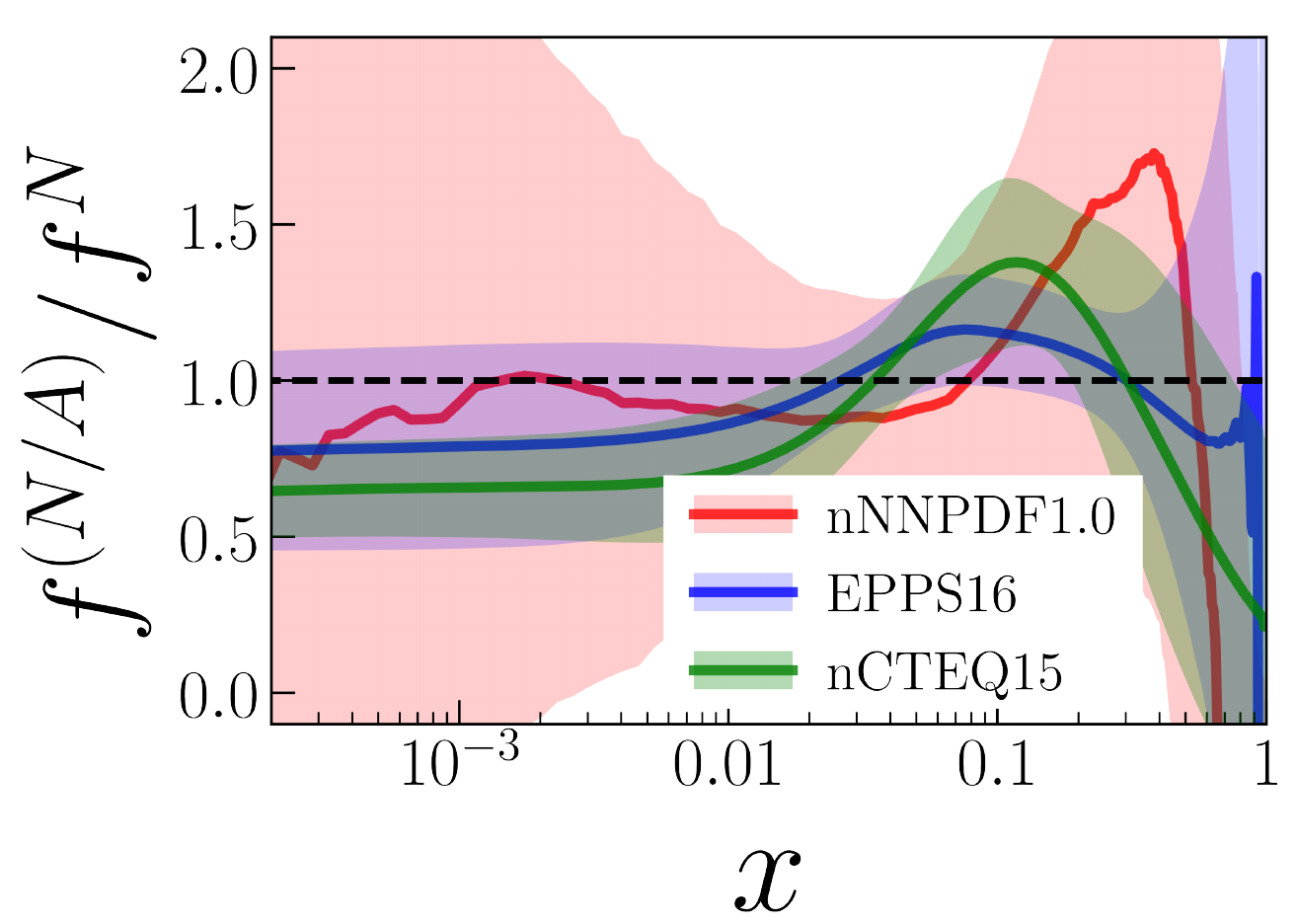}
\includegraphics[width=0.35\textwidth]{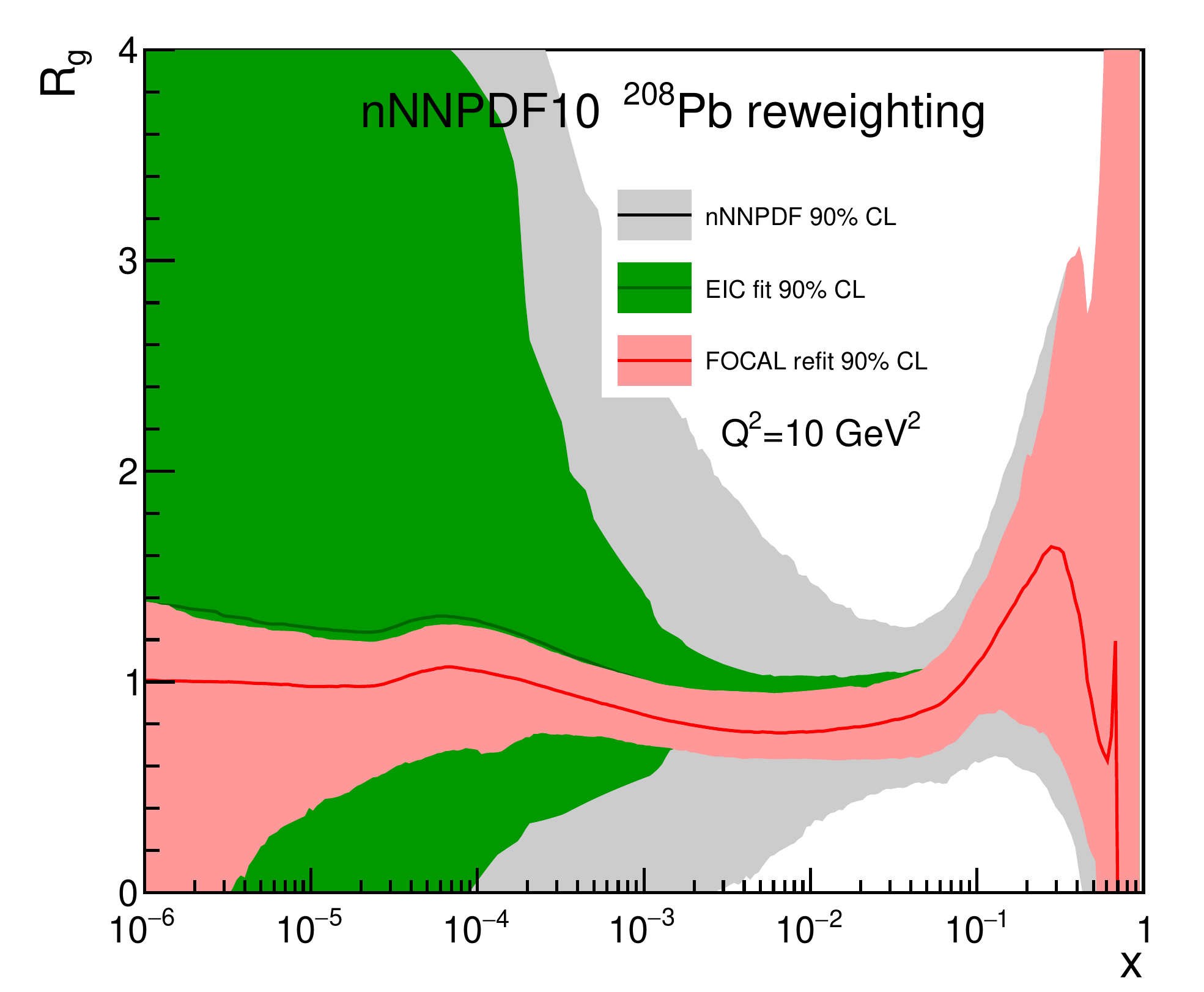}
\caption{Nuclear modification factor $R_g^{\mathrm{Pb}}$ of gluon PDFs for Pb nuclei at a scale of $Q^2 = 10 \, \mathrm{GeV}^2$ with their 90\% CL uncertainty bands. Left:  A comparison of $R_g^{\mathrm{Pb}}$ as obtained in fits of different theoretical frameworks to the same data.  (From \cite{nNNPDF}.) Right: Results using nNNPDFs with existing DIS data (grey band) and with an additional reweighting with data from future facilities: the EIC (green) and the FoCal detector in ALICE (light red). (From \cite{marco-pdf}).}
\label{fig-nnpdf}       % Give a unique label
\end{figure}

Recently, the NNPDF collaboration also provided a nuclear PDF set \cite{nNNPDF}. The neural net\-work approach allows for high flexibility, as it doesn't use a fixed analytical shape. This leads to even much larger uncertainties of nuclear PDFs, in particular at small $x$. The left panel of Fig.~\ref{fig-nnpdf} shows the nuclear modification factor of the gluon density in Pb as a function of $x$ for three different theoretical frameworks, when fitted to the same data \cite{nNNPDF}. Results for nCTEQ15 and EPPS16 show the same feature as mentioned before, i.e. no $x$-dependence at low x, and relatively narrow error bands. The fit result from the nNNPDF set shows significantly larger error bands. This demonstrates the impact of larger flexibility of the PDF parameterisation, and it strongly suggests that assumptions entering the nCTEQ15 and EPPS16 frameworks are too restrictive to allow for realistic error estimates.

While it is not clear, whether the recipe used by nNNPDF is fully justified, this seems to be a more promising approach. This has therefore been used in \cite{marco-pdf} for more recent estimates of the performance of future measurements related to nuclear PDFs as shown in the right panel of Fig.~\ref{fig-nnpdf}. Here the grey band again shows the uncertainties for a fit of nNNPDF to nuclear DIS data, where uncertainties are very large already for $x < 10^{-2}$. Using data of the future EIC facility, the uncertainties would be reduced to the green band, with strong constraints down to $x \approx 10^{-3}$ -- still, below this, uncertainties remain very large. A measurement of direct photons at forward rapidities at the LHC with the proposed FoCal detector would provide much stronger constraints still, significantly reducing uncertainties down to values as low as $x \approx 10^{-5}$.
\vspace{-2mm}

\vspace{-2mm}
 \section{Conclusion}
The small-$x$ region for proton PDFs is used by theorists to demonstrate the necessity of BFKL evolution, however, no proof of non-linear evolution has been observed. Nuclear PDFs with their much larger uncertainties should profit strongly from using LHC data as constraints. However, it is important to revisit the assumptions on small-$x$ behaviour in the theoretical descriptions to allow for realistic estimates of the uncertainty. Forward measurements at LHC, in particular those of real photons, have likely the best potential to provide information on the gluon density at very small $x$.

\small{\textit{(Material reproduced under CC-licence: https://creativecommons.org/licenses/by/4.0/.)}}

\end{document}